\documentclass
{article}
\usepackage[left=2cm,right=2cm,top=2cm,bottom=3.5cm,bindingoffset=0cm]{geometry}
\usepackage[T2A]{fontenc}
\usepackage[cp1251]{inputenc}
\usepackage{amsmath}
\usepackage{amssymb}
\usepackage{cite}
\usepackage{graphicx}
\usepackage{floatflt}
\usepackage{color}
\usepackage{epstopdf}
\DeclareGraphicsExtensions{.eps}
\newcommand{\bm}[1]{\mbox{\boldmath$#1$}}

\def\be{\begin{equation}}
\def\ee{\end{equation}}
\def\bea{\begin{eqnarray}}
\def\eea{\end{eqnarray}}
\def\nn{\nonumber}

\title{ON  THE  ``FORCE-FREE SURFACE '' OF THE MAGNETIZED CELESTIAL BODIES}
\author{V. Epp$^{a,b}$ and M. A. Masterova$^{a,c}$\\[10pt]
\itshape
$^{a}$Tomsk State Pedagogical University,  
ul. Kievskaya, 60,  634061  Tomsk, Russia\\
\itshape $^{b}$Tomsk State University 
 pr. Lenina, 36, 634050  Tomsk, Russia\\
\itshape
$ ^{c}$Tomsk Polytechnic University, pr. Lenina, 30, 634050 Tomsk, Russia}
\date{}
\begin{document}
\maketitle
\begin{abstract}
The field of a uniformly magnetized rotating sphere is studied with special attention to the surface where the electric and magnetic fields are orthogonal to each other.
The equation of this surface,  valid at arbitrary distances from the rotating magnetized sphere, is obtained. 
Inside the light cylinder this surface can be considered as a force-free surface, i.e. as a place where the  particles with strong radiation damping can be trapped due to their energy loss. Outside the light cylinder this surface makes just a geometric locus which moves with a superlight velocity around the axis of rotation. The 2- and 3-dimensional plots of the force-free surface are constructed. Estimation of influence of the centrifugal force on the particle dynamics is made. It is shown, that in case of strong magnetic field the centrifugal force is negligible small everywhere except a narrow neighbourhood of the light cylinder.
\end{abstract}

PACS: 94.30.-d; 97.60.-s; 41.20.-q

\section{Introduction}

Investigations of the magnetosphere of magnetized celestial bodies  is an important physical problem. The structure of the electromagnetic field, and dynamics of the charged particles in this field are the basic problems in understanding the nature of pulsars radiation (cf. \cite{Michel}).

The model of rotating neutron star with strong magnetic field was originally constructed by Goldreich and Julian \cite{Goldreich}. They suggested that the charged particles are moving along magnetic field lines and accelerated by the electric field component parallel to the magnetic field.  
 
It was shown in Ref. \cite{Jackson} that  over the polar caps of  rotating magnetized body with aligned magnetic and rotational axes there is a surface defined by equation $\bm E\cdot \bm H=0$. The electric field does not accelerate the charged particles  along the magnetic field lines in this region. By this reason this surface is referred to as a force-free surface. It was shown that the particle potential energy is minimal at the force-free surface. It was suggested that the particles which undergo the radiative damping force,  collect around this surface. If the axis of rotation  coincides with the magnetic axis,  the force-free surface forms a dome  above the polar cap which rests on the surface of the star.

The force-free surface in the vicinity of an inclined magnetic rotator was studied numerically in Refs \cite{Finkbeiner, Thielheim, Biltzinger}. The neutron star was considered as a homogeneously magnetized sphere rotating around an axes inclined with respect to the magnetic axes. It was shown by numerical integration of the particle trajectories that a relativistic  particle is captured at the $\bm E\cdot \bm H=0$ surface if the damping force is higher then some definite critical value \cite{Finkbeiner, Thielheim}. The perspective views of the force-free  surfaces for an oblique rotator were plotted using the vacuum fields of Deutsch \cite{Deutsch}. The force-free surface for a charged magnetized sphere was studied by Biltzinger and  Thielheim \cite{Biltzinger}.

The analytical expression for the force-free surface was obtained by Istomin and Sobyanin  \cite{Istomin} for the near-field region. It was shown that the radiation of charge particles  causes the trapping of these particles at the force-free surface. Further motion of the particles consists of relativistic oscillations near the surface and of  regular drift along the  surface. 
Some part of the charged particles captured by the force-free surface can move along the surface away from the centre of the field. As the azimuthal motion under the Coriolis force is restricted by the dipole magnetic field, the particles are moving with approximately constant angular velocity co-rotating with the neutron star magnetosphere. The kinetic energy of the particles increases due to the growth of the azimuthal velocity with respect to inertial reference frame. Thus accelerated charged particles leave the force-free surface in the vicinity of the light cylinder, maintaining the  pulsar wind \cite{Kirk, Kutschera}. In order to study this process, we have to know the geometry of the force-free surface not only in the vicinity of the neutron star, but throughout the entire space inside of the light cylinder.

In the following the force-free surface is considered at the distances up to the light cylinder and some remarks are made about the surface beyond the cylinder. In Section~\ref{sec2em} we obtain the equations of the force-free surface for the electromagnetic field of the nonrelativistic inclined magnetized sphere. The geometry of the force-free surface is studied in Section~\ref{sec3em}. We summarize our results and discuss the further possible developments in Section~\ref{sec4em}.

A few words about notations. Greek letters are reserved to label spacetime indices and run the set of values $0,1,2,3$ while Latin letters correspond to only spatial indices $1,2,3$. Throughout the work summation over repeated indices is implied.
\section{Equation of the force-free surface}\label{sec2em}
The electromagnetic field of a neutron star can be approximated by the field of inclined rotating magnetized sphere  \cite{Michel,Deutsch,Babcock}. Taking into account that a solid sphere is incompatible with the theory of relativity (see related discussion in \cite{EppM2013, EppM2014} and references therein) we consider the electromagnetic field of a nonrelativistic rotating body. This field is described by the 4-vector potential $A^{\nu}$, which in the coordinate system with temporal $t$ and spherical $r,\theta,\varphi$ coordinates, has the following components \cite{EppM2014} (axis $Z$ is directed along the vector of angular velocity)
\bea\label{potent}
\begin{aligned}
&A^0=-\frac{\omega r_0^2\mu}{6 c r^3}\left[3C\sin2\theta\sin\alpha+\cos\alpha(3\cos2\theta+1)\right],\\
&A^1=0,\quad A^2=-\frac{\mu}{r^3}S\sin \alpha,\quad A^3=\frac{\mu}{r^3\sin\theta}(\cos\alpha\sin\theta-C\sin \alpha\cos\theta),
\end{aligned}
\eea
where  $S=\sin\lambda-\rho\cos\lambda$, $C=\cos\lambda+\rho\sin\lambda$, $\lambda=\varphi-\omega t+\rho$, $\rho=r\omega/c$, $\mu$ is the module of the dipole moment vector of the magnetized sphere rotating with angular frequency $\omega$ and forming an angle $\alpha$  between the dipole  vector and axis $Z$, $r_0$ is the radius of the sphere, and $c$ is the speed of light. The field described by the potential  (\ref{potent}) coinsides with the field of Deutsch \cite{Deutsch} in approximation of a nonrelativistic sphere $(r_0\omega/c\ll 1)$.

Let us find the equation of the surface specified by condition $\bm E\cdot\bm H=0$. Note that time and the azimuthal angle $\varphi$ are represented  in the formula for potential  only in combination $\phi -\omega t$. That means that the equation of the force-free  surface in the co-rotating reference frame does not depend on time. On the other hand, the equation $\bm E\cdot\bm H=0$ can be written in a covariant form $F^{\mu\nu}F_{\mu\nu}= 0$,  where  $F_{\mu\nu}=\partial_{\mu}A_{\nu}-\partial_{\nu} A_{\mu}$ is the electromagnetic field tensor. Hence, the shape of the surface does not  depend on choice of the  reference frame. The derivation of the  of force-free surface equation in the rotating reference frame is slightly more complicated and involves the  issues of definition of the vectors $\bm E$ and $\bm H$ in the non-inertial reference frame. We shall develop this equation in the co-rotating reference frame, just to show how it works. 
Besides, this reference frame is more convenient for numerical calculations and graphical illustrations which we present in the next sections 3 and 4.

We define the coordinates  in the co-rotating  reference frame as follows: $x^{0'}=ct, x^{1'}=r, x^{2'}=\theta, x^{3'}=\psi=\varphi-\omega t$. The vector potential  (\ref{potent}) in this frame  reads:
\bea\label{potent22}
\begin{aligned}
& A_{0'}=\frac{\mu\omega}{cr}\left[C\frac{\sin2\theta }{2}\sin\alpha\left(1-\frac{r_{0}^2}{r^2}\right)-\sin^2\theta \cos\alpha\left(1-\frac{r_{0}^2}{r^2}\right)-\frac{2}{3}\cos\alpha \frac{r_{0}^2}{r^2}\right],\\
& A_{1'}=0,\qquad A_{2'}=\frac{\mu}{r}S\sin \alpha,\qquad A_{3'}=\frac{\mu\sin\theta}{r}(C\sin\alpha\cos\theta-\sin\theta\cos\alpha).
\end{aligned}
\eea
This gives the following components of  electromagnetic field tensor $F_{\mu'\nu'}=\partial_{\mu'} A_{\nu'}-\partial_{\nu'} A_{\mu'}$:
\begin{align}
& F_{0'1'}=-\frac{3\omega r_{0}^2 \mu }{2 c r^3}\left[\sin2\theta \sin\alpha \left(C-\rho ^2\frac{\cos\lambda}{3}\right)+\cos\alpha (3\cos2\theta +1)\right]-\nn\\
&\qquad\quad\,-\frac{\mu\omega }{cr^2} \left[\sin^2\theta\cos\alpha-\frac{\sin2\theta}{2}\sin\alpha (C-\rho^2\cos\lambda )\right],\nn\\[5pt]
\label{F}
&F_{0'2'}=\frac{\mu \omega }{c r}\left[\frac{r_{0}^2}{r^2}(C\cos2\theta \sin \alpha-\sin2\theta \cos\alpha)-C\cos 2\theta\sin\alpha+\sin2\theta\cos\alpha \right],\\[5pt]
&F_{0'3'}=\frac{\mu\omega}{2cr} S\sin 2\theta\sin\alpha\left(1-\frac{r_{0}^2}{r^2}\right),\, F_{1'3'}=\frac{\mu}{r^2} \sin\theta\left[\sin\theta\cos\alpha-\cos\theta\sin\alpha (C-\rho^2\cos\lambda )\right],\nn\\[5pt]
&F_{1'2'}=-\frac{\mu}{r^2} \sin \alpha \left(S-\rho^2\sin^2\lambda\right),\quad\, F_{2'3'}=-\frac{2\mu}{r} \sin\theta(C\sin\theta\sin\alpha+\cos\theta\cos\alpha).\nn
\end{align}
The 3-vectors of the electric field  $E_i$ and  of the magnetic  field $H_i$ in arbitrary  curved spacetime are defined by expressions \cite{Landau, Moller} (further on, the primes are omitted)
\bea\label{111}
E_i =F_{0\,i}, \qquad B^i =-\frac{1}{2\sqrt{\gamma}}\varepsilon^{ijk} F_{jk},
\eea
where $\varepsilon_{ijk}$ is the completely antisymmetric Levi-Civita tensor, $\gamma$ is the determinant of the space metric tensor
\bea
\gamma _{ij}=-g_{ij}+\frac{g_{0i }g_{0j}}{g_{00}}.
\nn
\eea

The force-free surface is defined by equation $E_i B^i =0$. Substituting the components $F_{\mu'\nu'}$ from   (\ref{F}) we obtain
\bea\label{FFSer}
\begin{aligned}
&4\cos\theta \Big\{\cos\theta \cos\alpha +\sin\theta \sin\alpha \big[\cos(\psi +\rho)+\rho \sin(\psi +\rho)\big]\Big\}^2+\\
&+\sin\alpha \Big\{\big[\cos(\psi +\rho)+\rho \sin(\psi +\rho)\big]\cos\alpha \sin\theta -\sin\alpha \cos\theta\Big\}\left(\frac{\rho^2}{a^2}-1\right)=0,
\end{aligned}
\eea
where the notation  $a=\omega r_0/c$ is introduced. The equation (\ref{FFSer}) defines the force-free surface in the rotating reference frame  for the inner area of the light cylinder, i.e. for $\rho < 1$.

In case of $\rho\ll1$ this yields
\bea\label{pot2}
\rho ^2=a^2\left(1-4\frac{\cos\theta [\cos\theta \cos\alpha +\sin\theta \sin\alpha\cos\psi ]^2}{\sin\alpha [\cos\psi\cos\alpha \sin\theta -\sin\alpha \cos\theta ]} \right).
\eea
This expression was originally obtained in paper \cite{Istomin}. As one can see, the geometry of the force-free surface does not depend on $a$ in this approximation. The radius of the sphere plays a role of a scale factor. However, if $\rho$ is comparable with unity, the force-free surface has more complicated form. In particular, it is wound up around the $Z$-axis due to the argument $\psi +\rho$ of the trigonometric functions in equation (\ref{FFSer}).

One can easily check that equation of the  force-free surface in the inertial reference frame is defined by the same equation (\ref{FFSer})  except the substitution $\psi=\varphi-\omega t$.
 However, in contrast to  equation  (\ref{FFSer}), the equation in the rest reference frame
 holds both  in the inner space of the  light cylinder and beyond the light cylinder\footnote{In this connection a question arises -- how  the force-free surface behaves in the far-field zone?
 As is well-known the vector $\bm{E}$ is orthogonal to the vector $\bm{H}$ everywhere in this zone and, consequently, the force-free surface has to fill all the space.Can the two-dimensional surface transform into the three-dimensional space? Answers to these questions are given in Appendix.}.

\section{Geometry of the force-free surface}\label{sec3em}
 The value of $a$ in Eq. (\ref{FFSer}) depends sufficiently on the angular velocity of a celestial body. The greatest values  of $a$ belong  presumably to pulsars. For example, $a\approx0.2$ for the pulsar PSR J1748-2446ad. We have plotted the views of the force-free surface for different values of radius $a$ and the inclination angle $\alpha$ using the equation (\ref{FFSer}).
In case of the aligned rotator ($\alpha=0$), the equation of the force-free surface reads: $\cos\theta=0$. This equation holds in the total equatorial plane $\theta=\pi/2$.

\begin{figure}[htbp]
\begin{center}
\includegraphics[width=2in]{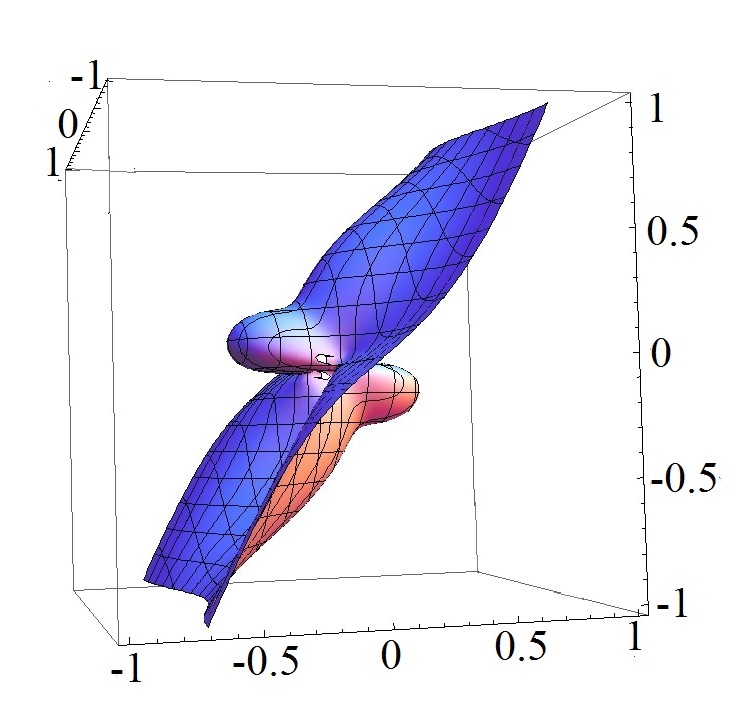}
\hspace{2cm}
\includegraphics[width=2in]{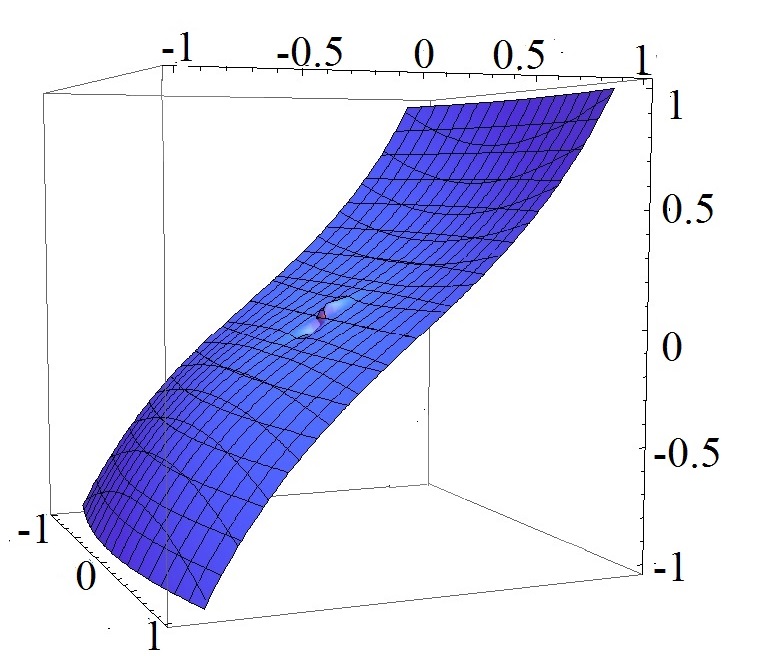}
\end{center}
\caption{The force-free surface for $\alpha =\frac{\pi }{3}$; $a^2 =0.1$ (left) and $a^2 =0.001$ (right)}
\label{fig8-3d}
\end{figure}

\begin{figure}[htbp]
\begin{center}
\includegraphics[width=2.0in]{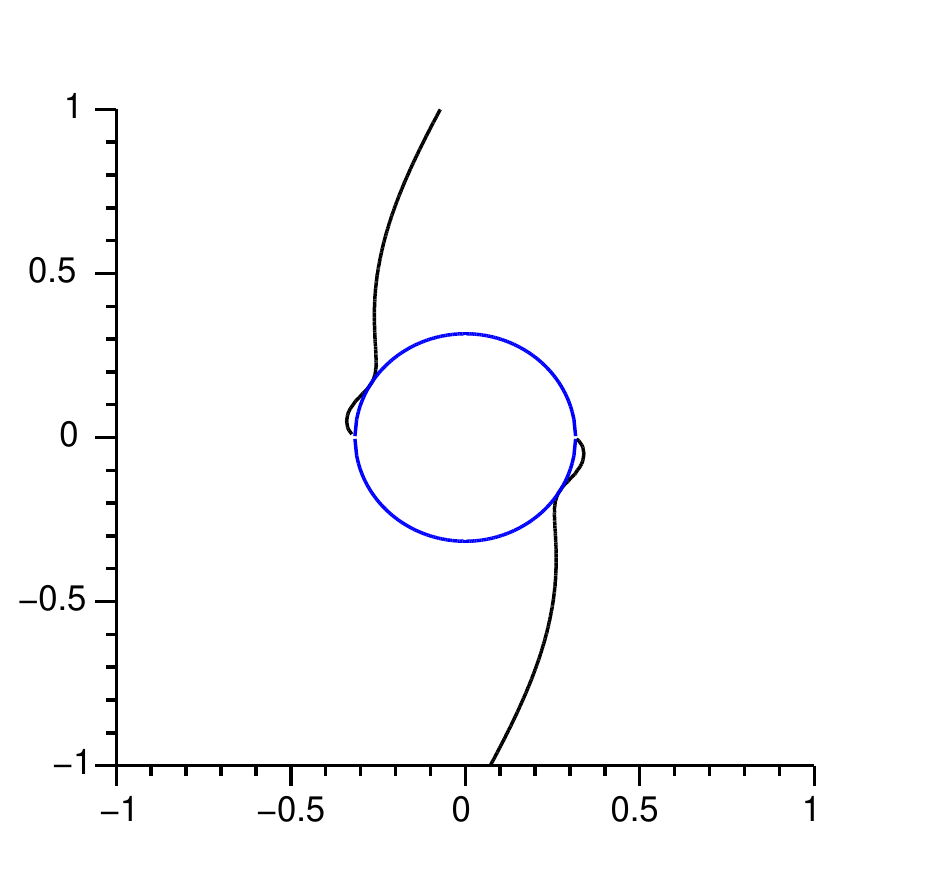}
\includegraphics[width=2.0in]{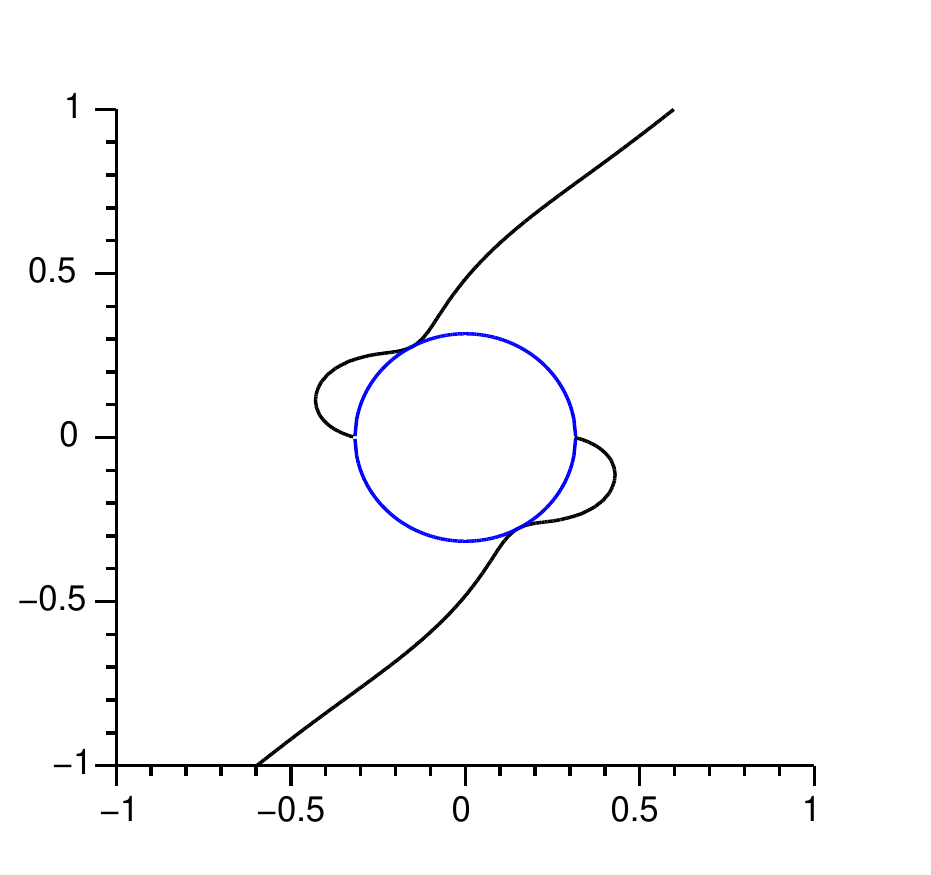}
\includegraphics[width=2.0in]{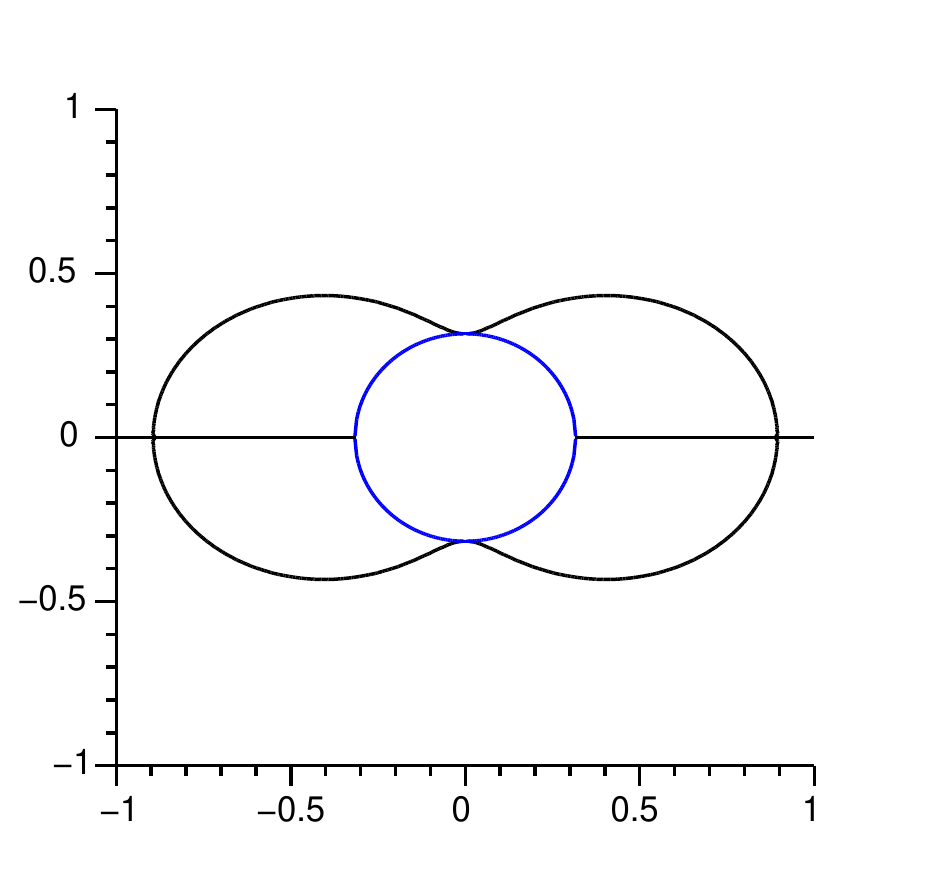}\\
\includegraphics[width=2.0in]{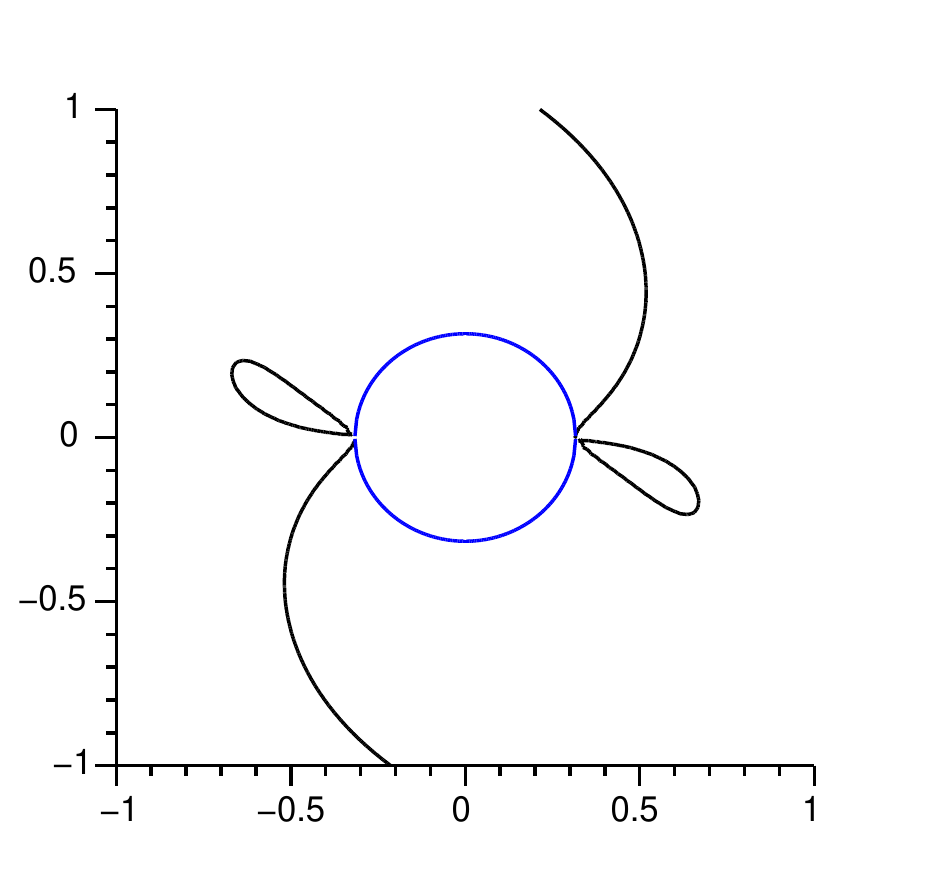}
\includegraphics[width=2.0in]{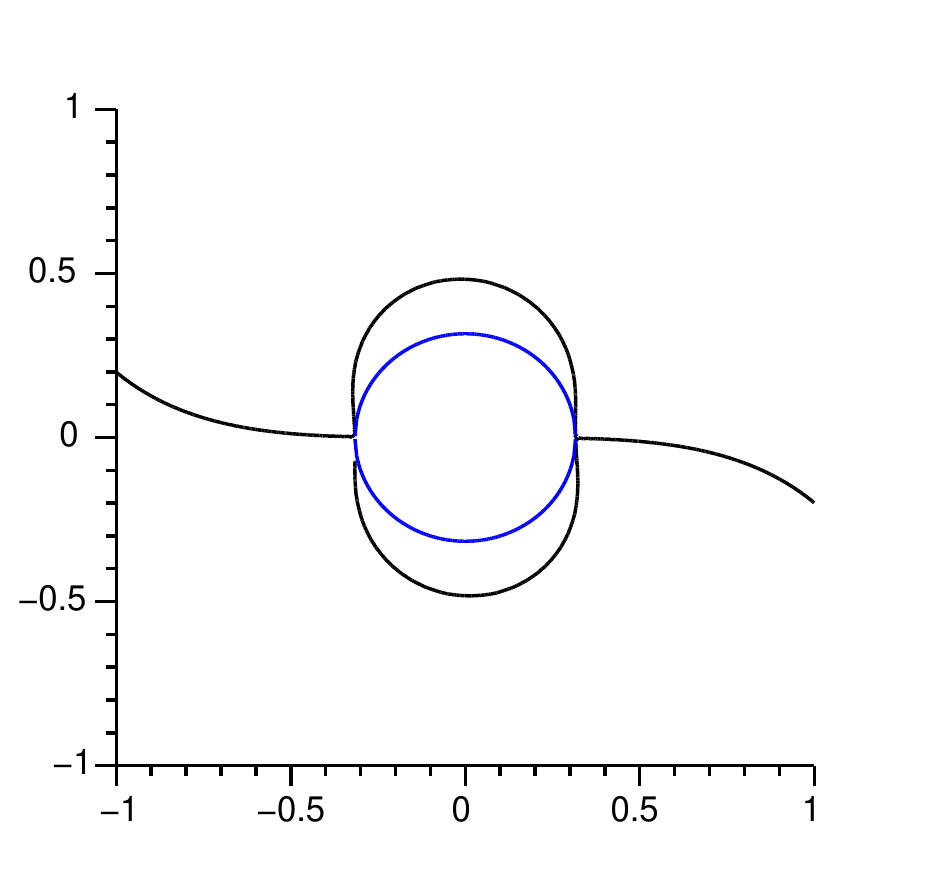}
\includegraphics[width=2.0in]{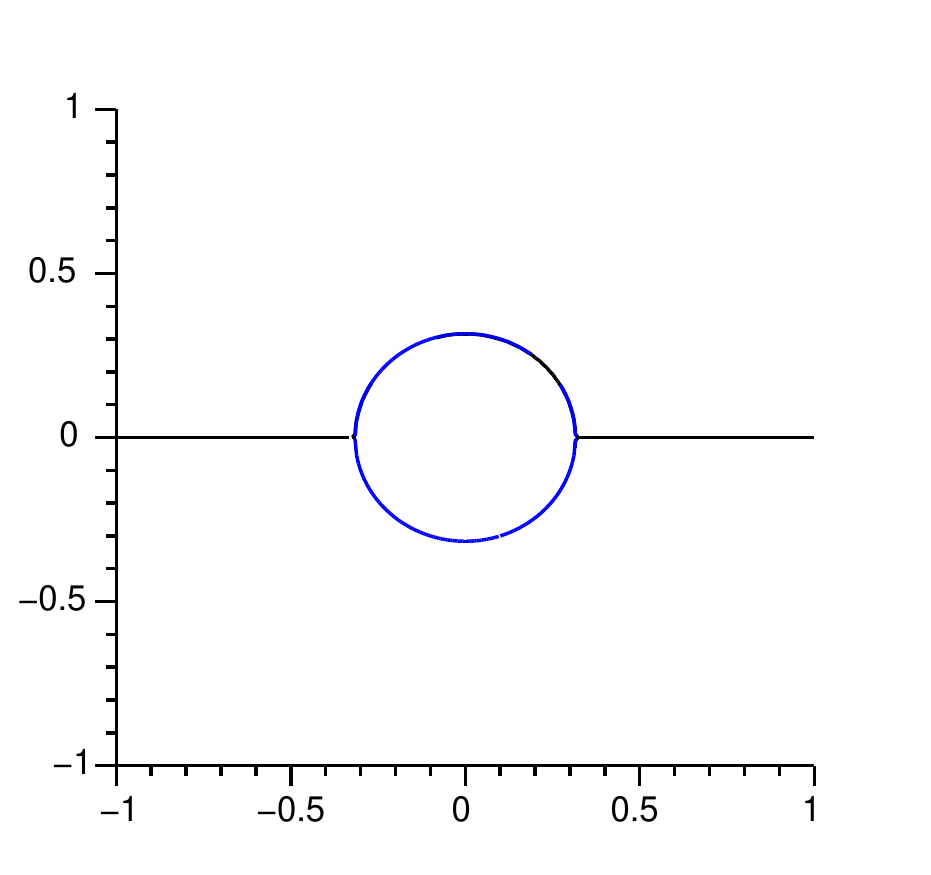}
\end{center}
\caption{Sectional views of the force-free surface for $a^2 =0.1$ in the plane $ \psi ={0,\pi}$ (first row), and in the plane
$\psi =\displaystyle\frac{\pi}{2},\displaystyle\frac{3\pi}{2}$ (second row);  $\alpha= \displaystyle\frac{\pi }{6},\frac{\pi }{3},\frac{\pi }{2}$, with   $\alpha$ increasing from  left to right}
\label{figFFS0-1}
\end{figure}

\begin{figure}[h!]
\begin{center}
\includegraphics[width=2.0in]{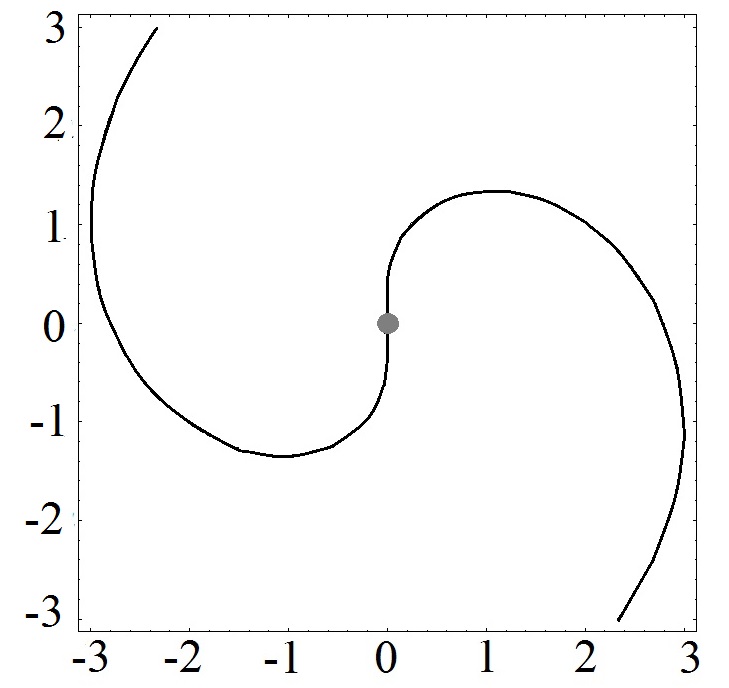}
\caption{Sectional view of the force-free surface in the plane $\theta=\pi/2$}
\label{fig2em}
\end{center}
\end{figure}

\begin{figure}[htbp]
\begin{center}
\includegraphics[width=2in]{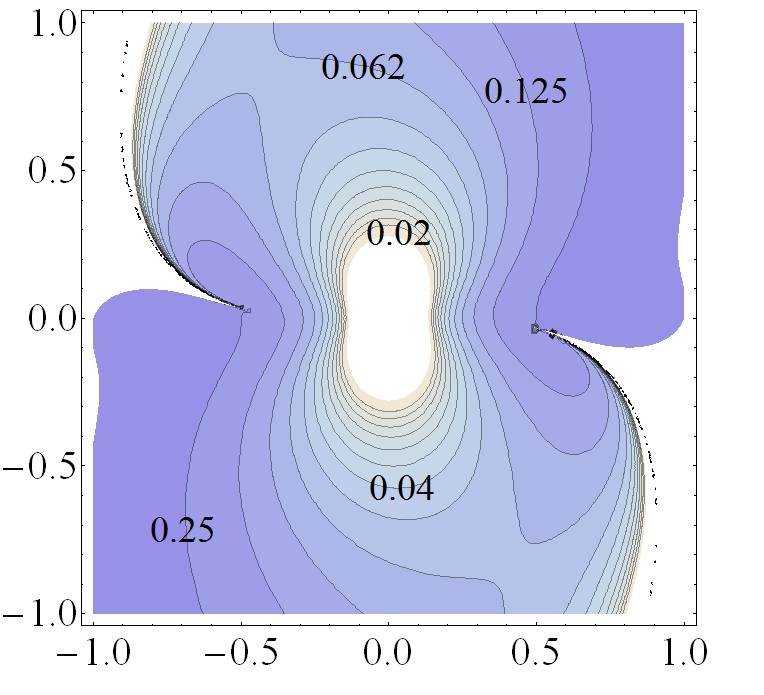}
\hspace{2cm}
\includegraphics[width=2in]{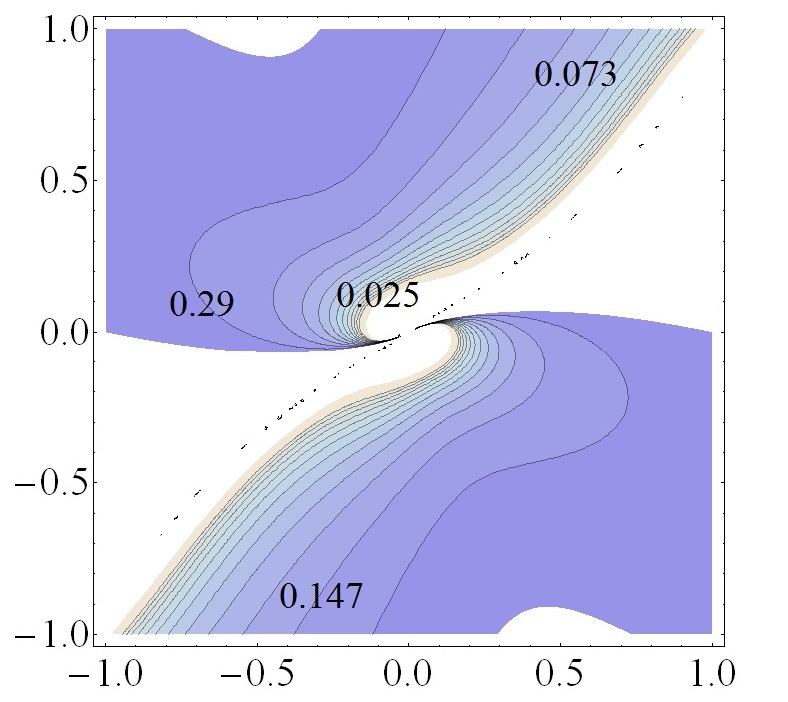}
\caption{Sectional view of the force-free surface for $\alpha=\displaystyle\frac{\pi }{6}$ in the plane
$\psi ={\displaystyle\frac{\pi}{2},\displaystyle\frac{3\pi}{2}}$ (left), and in the plane $\psi ={0,\pi}$ (right) for different values of the radius of the sphere $a$. The values of $a$ are marked on the figures.}
\label{fig1-dif}
\end{center}
\end{figure}

The structure of the force-free surface for $a^2=0.1$ and $a^2=0.001$, and for angle $\alpha=\pi/3$ is shown in Fig. \ref{fig8-3d}.
The sectional views of the surface    $\bm E\cdot \bm H=0$ for $a^2=0.1$ and different values of angle $\alpha$  are represented in Fig. \ref{figFFS0-1}. 
The plots in the first row  look
similar to those drawn in Refs  \cite{Finkbeiner, Istomin} for small $\rho$, but figures in the second row differ essentially,  especially for $\alpha=\pi/6$.
This difference is caused by the fact that in case of $\rho$ comparable with the unity, one has to take into account  the variable $\rho$ in  argument of the trigonometric functions in Eq. (\ref{FFSer}). Due to this argument, the force-free surface becomes  twisted around the $Z$-axis. This is clearly seen in the equatorial sectional view in Fig. \ref{fig2em}.
The curve in this cutting section is determined by equation
\[
\big[\cos(\varphi-\omega t +\rho)+\rho \sin(\varphi-\omega t +\rho)\big]\left(\frac{\rho^2}{a^2}-1\right)=0
\]
for a fixed instant of $t$. Its shape does not depend on the inclination angle $\alpha$. The surface $\bm E\cdot \bm H=0$ is represented  by  the circle of radius $a$ and the curve outside this circle.
The dependence of the force-free surface on $a$ is graphically represented on the Fig. \ref{fig1-dif} as sectional views for $0.02<a<0.3$.

\section{The centrifugal force}\label{sec4em}
The force-free surface is of interest  because the  particles under strong radiation damping
can be trapped in regions where $\bm E\cdot\bm H=0$. Making this statement we do not take into account that the particles, co-rotating with the star, undergo the centrifugal force. In the near region of an oblique rotator and in case of strong magnetic field the centrifugal force can be neglected. However, in the regions near the light cylinder the centrifugal force affects sufficiently the particle dynamics. Hence, the surface $\bm E\cdot\bm H=0$ is no longer a force-free one.

Let us estimate the distance at which the  centrifugal force should be taken into account.
It follows from the relativistic Hamiltonian function in the rotating reference frame that he potential of the centrifugal force is equal to
$U=mc^2\sqrt{1-R^2}$, where $R=\rho\sin\theta$ is the distance from the axis of rotation in units of the light cylinder radius. The module of the centrifugal force in dimensionless representation reads
\begin{eqnarray}
F_{c}=\frac{R}{\sqrt{1-R^2}}
\end{eqnarray}
The longitudinal with respect to $\bm H$ component of the electric field acts on the particle by force
\begin{eqnarray}
\label{P-1}
F_{e}=e E_{H}=e\frac{(\bm{E}\bm{H})}{H},
\end{eqnarray}
which in the same dimensionless representation has the form
\be
\label{F-field-N}
F_{e}\sim\frac N\rho,\quad N=\displaystyle\frac{e\mu\omega^2}{m c^4}.
\ee
Then  the centrifugal force and the force of electromagnetic interaction are in the ratio
\be
\frac{F_c}{F_e}\sim\frac{R\rho}{N\sqrt{1-R^2}}.
\ee
It is readily seen
 that in case of small $N$ the centrifugal force can be neglected if $R\ll N$. But in  case of large $N$ (which is typical for the neutron stars) this force can be neglected everywhere with exception of the small vicinity of the light cylinder: $1-R\sim N^{-1}$.

\section{Conclusions}

We have obtained  in the preceding sections the equation for the surface $\bm E\cdot\bm H=0$  valid at arbitrary distances from the rotating magnetized sphere. This surface co-rotates with the magnetized sphere as a unit.
Inside the light cylinder this surface can be considered as a force-free surface, i.e. as a place where the  particles with strong radiation damping can be trapped due to their energy loss as described in Refs  \cite{Finkbeiner, Biltzinger, Istomin}. Outside the light cylinder the surface $\bm E\cdot\bm H=0$ is just a geometric locus which moves with superlight velocity (see Appendix).

The main purpose of this paper is to show that in case of highly magnetized sphere ($N\gg 1$, with $N$ defined by Eq. (\ref{F-field-N})), the surface  $\bm E\cdot\bm H=0$ can be used in study of the particles dynamics in the entire space within the light cylinder, not only in the region $\rho\ll 1$. In this case one has to take into account that the force-free surface is twisted around the rotational axis, if we consider the distances $\rho\sim 1$. The only exception is the thin belt at the inner side of the light cylinder ($1-N^{-1}\lesssim R<1$) where the centrifugal force is comparable or greater than the force of electromagnetic interaction.
In case of weak electromagnetic field ($N\ll 1$) the surface $\bm E\cdot\bm H=0$ can be considered as a force-free one only in the region $R\ll N$. Otherwise, the centrifugal force should be taken into account.

The particles captured by the force-free surface inside the light cylinder are leaving this surface in the vicinity of the light cylinder, forming an outward flow of relativistic particles. Since the force-free surface co-rotates with the neutron star, this flow of particles has the shape of a sequence of expanding quasi-spherical surfaces having the equatorial section shown in Fig. \ref{fig2em}. This flow of particles resembles the pulsar wind, described in details in \cite{Kirk}. Of course, the results of this paper are not applicable to magnetosphere filled with dense plasma.

\section*{Acknowledgements}
The work was supported by Ministry of Education and Scence of Russian
Federation, project No 867.
\section*{Appendix}
Although the surface $\bm E\cdot \bm H=0$ makes no sense in study the particle dynamics in the far-field region, it is interesting from the academic point of view to investigate the asymptotic $\rho\to\infty$.

The scalar product $\bm E \cdot \bm H$ obtained from the Deutsch field \cite{Deutsch} in approximation $\rho\gg 1$ is, evidently, equal to zero because this field corresponds to radiation.
In order to find the equation of force-free surface when $\rho\gg 1$ we consider the equation (\ref{FFSer}) in inertial reference frame. If we take into account only the components which are proportional to the highest power of $\rho$, then we obtain the following equation
\[
\cos\alpha \sin\theta\sin\alpha \sin(\varphi-\omega t +\rho)=0.
\]
If $\alpha\neq 0,\pi/2,\pi$ and $\theta\neq 0,\pi$ we have the equation of a spiral which rotates with the angular velocity $\omega$ around the  $Z$--axis
\be\label{spiral}
\rho=\omega t-\varphi.
\ee
The pitch of the spiral is equal to $\Delta\rho=2\pi$ or, in dimensional units, $\Delta r=2\pi c/\omega$ which is equal to the wave length of dipole radiation. As we see, the force-free surface has a form of a surface which wounds on the $Z$--axis  in such a way that the line of its intersection with an arbitrary cone $\theta={\rm const}$ is a spiral described by the equation (\ref{spiral}). Since the equation (\ref{spiral}) does not depend on the angle $\theta$, each turn of this surface is close to a sphere. Rotation of the such quasi-spherical spiral surface produces a picture of expanding set of concentric spheres, each of them can be considered  as a wave-front of a radiation field.

Hence, the ``force-free'' surface does not fill all the space in the wave zone. Strictly speaking, the electric field is not orthogonal to magnetic one in between the adjacent windings of the surface $\bm E\cdot \bm H=0$ even in the far field region. Claiming that $\bm E\cdot\bm H=0$ in the wave zone, we are neglecting the terms in  electromagnetic field equations, which decrease faster than $1/r$. If we take into account these terms, we shall see that there is a component of $\bm E$ parallel to $\bm H$, but it is of order  $\lambda/r$, where $\lambda$ is the wave length of the dipole radiation.


\begin{thebibliography}{nn}
\bibitem
{Michel}  F.C. Michel, {\sl Theory of Neutron Star Magnetospheres.}  The University of Chicago Press: Chicago and London 1991.
\bibitem
{Goldreich} P. Goldreich, W.H. Julian, {\it ApJ} {\bf 157}, 869 (1969).
\bibitem
{Jackson} E.A. Jackson, {\it ApJ} {\bf 206}, 831 (1976).
\bibitem
{Finkbeiner}B. Finkbeiner, H. Herold, T. Ertl, and H. Ruder,
 {\it Astron. Astrophys} {\bf 225}, 479 (1989).

\bibitem{Thielheim} K.O.~Thielheim, H.~Wolfsteller,
{\it ApJ Suppl. Ser.} {\bm 71}, 583 (1989).

\bibitem{Biltzinger}P. Biltzinger and K.O. Thielheim,
arXiv:astro-ph/0011306 (2004).

\bibitem
{Deutsch} A.J. Deutsch, {\it Ann. d'Astrophys.}  {\bf 18}, 1 (1955).
\bibitem
{Istomin} Ya.N. Istomin, D.N. Sobyanin,
{\it J. Exp. Theor. Phys.}, {\bf 109}, 393 
 (2009).

\bibitem{Kirk}J.G. Kirk, Y. Lyubarsky, J. Petri, in: 
{\it Neutron Stars and Pulsars}, Astrophys. and Space Sci. Lib. {\bf 357} 421 (2009).

\bibitem{Kutschera}
 M.~Kutschera, D.~Gora, P.~Homola, J.~Niemiec, B.~Wilczynska, H.~Wilczynski,
 {\it Acta Phys. Pol.} {\bf B 35} 1837  (2004)

\bibitem{Babcock}
H.W. Babcock, T.G. Cowling, {\it Mon. Not. R. Astron. Soc.}, {\bf 113}, 356 (1953).
\bibitem%
 {EppM2013} V. Epp, M.A. Masterova, {\it Astrophys. Space Sci.} {\bf 345}, 315 (2013).
\bibitem%
 {EppM2014} V. Epp, M.A. Masterova, {\it Astrophys. Space Sci.} {\bf 353}, 473 (2014).
 \bibitem
{Landau}L.D.~Landau, E.M.~ Lifshitz, {\it The Classical Theory of Fields}, Pergamon, N.Y. 1975.
\bibitem
{Moller} C. M{\o}ller, {\it The Theory of Relativity}, Oxford University Press, Oxford 1976.
\end{thebibliography}
\end{document}